\documentclass[lettersize,journal]{IEEEtran}
\usepackage{amsmath,amsfonts}
\usepackage{algorithmic}
\usepackage{algorithm}
\usepackage{array}
\usepackage[caption=false,font=footnotesize,labelfont=sf,textfont=sf]{subfig}
\usepackage{textcomp}
\usepackage{stfloats}
\usepackage{url}
\usepackage{verbatim}
\usepackage{graphicx}
\usepackage{cite}
\usepackage{colortbl}
\usepackage{bm}
\usepackage{fix-cm}
\usepackage{cite}
\usepackage{amsmath,amssymb,amsfonts}
\usepackage{graphicx,color}
\usepackage{booktabs}
\usepackage[colorlinks, linkcolor=red, anchorcolor=blue, citecolor=blue]{hyperref}
\usepackage{balance}
\usepackage{orcidlink}
\usepackage{multirow}
\usepackage{bbding}
\usepackage{makecell}
\usepackage{enumitem}
\usepackage{amsthm}

\newtheorem{remark}{Remark}

\begin{document}

\title{Dual Fluid Antenna-Assisted UAV MIMO Networks}

\author{
\IEEEauthorblockN{
Runke Fan,
Tianheng Xu, \IEEEmembership{Member, IEEE,}
Pei Peng, \IEEEmembership{Member, IEEE,}
Xianfu Chen, \IEEEmembership{Senior Member, IEEE,}
}
\IEEEauthorblockN{
Celimuge Wu, \IEEEmembership{Senior Member, IEEE,}
Kai-Kit Wong, \IEEEmembership{Fellow, IEEE,} and 
Mohsen Guizani, \IEEEmembership{Fellow, IEEE}}

\vspace{-2ex}
\thanks{R. Fan is with the Shanghai Advanced Research Institute, Chinese Academy of Sciences, Shanghai 201210, China, and is also with the University of Chinese Academy of Sciences, Beijing 100049, China (e-mail: fan-runke2025@sari.ac.cn).}
\thanks{T. Xu is with the Shanghai Advanced Research Institute, Chinese Academy of Sciences, Shanghai 201210, China (e-mail: xuth @sari.ac.cn).}
\thanks{P. Peng is with the School of Telecommunication and Information Engineering, Nanjing University of Posts and Telecommunications, Nanjing 210003, China (e-mail: pei.peng@njupt.edu.cn).}
\thanks{X. Chen is with the Shenzhen CyberAray Network Technology Co., Ltd., Shenzhen 518038, China, and is also with the Shanghai Advanced Research Institute, Chinese Academy of Sciences, Shanghai 201210, China (e-mail:xianfu.chen @ieee.org).}
\thanks{C. Wu is with the Graduate School of Informatics and Engineering, The University of Electro-Communications, Tokyo 1828585, Japan (e-mail: celimuge@uec.ac.jp).}
\thanks{
K. K. Wong is with the Department of Electronic and Electrical Engineering, University College London, Torrington Place, WC1E 7JE, United Kingdom and is also with the Department of Electronic Engineering, Kyung Hee University, Yongin-si, Gyeonggi-do 17104, Korea (e-mail: kai-kit.wong@ucl.ac.uk).}
\thanks{
M. Guizani is with Machine Learning Department, Mohamed
Bin Zayed University of Artificial Intelligence, Abu Dhabi, UAE (email: mohsen.guizani@mbzuai.ac.ae).}
\thanks{Corresponding author: Tianheng Xu.}
}

\markboth{IEEE Internet of Things Journal,~Vol.~14, No.~8, January~2026}%
{Shell \MakeLowercase{\textit{et al.}}: A Sample Article Using IEEEtran.cls for IEEE Journals}


\maketitle

\begin{abstract}
Fluid Antennas (FAs)-assisted Unmanned Aerial Vehicle (UAV) networks leverage the FA position adaptivity and flexible beamforming to overcome the limitations of Fixed-Positioned Antennas (FPAs) in dynamic UAV channels and Multi-User (MU) interference. This letter investigates a dual FA-assisted UAV network for MU-Multiple-Input-Multiple-Output (MIMO) downlink communications, aiming to maximize the average achievable rate through the joint optimization of UAV trajectory, the transmit/receive FA positions, and beamforming. The formulated problem is highly coupled and non-convex. Accordingly, an efficient Alternating Optimization (AO)-based algorithm is developed for decomposed subproblems, yielding a suboptimal solution. Numerical results demonstrate significant performance gains of 120$\textbf{\%}$ and 110$\textbf{\%}$ over conventional FPA-based and existing FA-based baselines, respectively.
\end{abstract}


\section{Introduction}

\IEEEPARstart{W}{ith} the advancement of Sixth-Generation (6G) technologies, ubiquitous connectivity has become a fundamental vision. In this context, Unmanned Aerial Vehicle (UAV) networks have emerged as a key technology for their high flexibility and rapid deployment to complement the terrestrial coverage. However, constrained by Fixed-Positioned Antenna (FPA) arrays, UAV networks still face critical challenges in dynamic propagation environments \cite{n1}. Specifically, Three-Dimensional (3D) UAV mobility induces channel instability and phase sensitivity, while energy constraints and Multi-User (MU) interference restrict the system throughput.

Fluid Antenna (FA) is considered a promising solution to address the above challenges, where the core idea is to adjust the antenna positions, known as ports, and rotation angles based on real-time channel conditions \cite{n2}. Compared with FPAs, FAs introduce additional spatial Degrees of Freedom (DoFs) and spatial diversity gain to mitigate channel variations and dense interference in UAV networks, thus enhancing the system performance.

The effectiveness of jointly optimizing FA positions and the UAV deployment has been verified in recent studies \cite{1, 2, 3, 4}. 
In \cite{1}, Tang et al. improved the phase-sensitive beamforming gains for MU interference mitigation by jointly adjusting FA positions, transmit beamforming, and UAV deployment.
In \cite{2}, Mao et al. proposed a robust FA-assisted UAV framework to address the uncertainty in the angle of departure induced by UAV jitter.
In \cite{3}, Reda et al. coordinated FA positions with phase shifts of the
RIS to enhance interference mitigation in cellular-connected UAV systems.
In \cite{4}, Ghadi et al. derived a compact analytical expression for the outage probability of UAV-relaying networks with FAs deployed at terrestrial users. 
Furthermore, the UAV 3D trajectory may lead to more complex channel conditions, and algorithm designs \cite{5, 6}. 
In \cite{5}, Zhang et al. proposed a uplink data collection framework for FA-assisted UAVs, where the sum rate is maximized through joint optimization of UAV trajectory, beamforming, power allocation, and FA positioning. 
In downlink transmission, Liu et al. in \cite{6} utilized an FA-assisted UAV to maximize the sum rate. The FA positions, transmit beamforming, and the UAV trajectory are alternatively optimized.
However, most existing studies depend on simplified Line of Sight (LoS) channels, neglecting the phase sensitivity under UAV mobility. Moreover, current joint designs of UAV trajectories and FAs are restricted to single-side FAs, failing to fully utilize the spatial DoFs of FAs.

In this work, we propose an Alternating Optimization (AO) algorithm for a dual-FA-assisted UAV MU-Multiple-Input-Multiple-Output (MIMO) network. The main contributions are listed as follows.
\begin{itemize}
    \item We introduce a general FA-assisted UAV MU-MIMO downlink channel model with a Two-Dimensional (2D) FA at the UAV and One-Dimensional (1D) FAs at terrestrial users, explicitly capturing the practical phase-related features, such as UAV yaw and Doppler shifts.
    
    \item We propose an efficient AO algorithm to maximize the average achievable rate. The non-convex problem is decomposed into tractable subproblems of the UAV trajectory, transmit/receive FA positions, and beamforming.

    \item Numerical results demonstrate the performance gains of the proposed algorithm over traditional the FPA-based baseline, with more stable convergence and higher rates than the state-of-the-art FA-based baseline.
\end{itemize}

\section{System Model and Problem Formulation}
We consider an FA-assisted UAV networks where a UAV 

\noindent serves $K$ terrestrial users. The UAV utilizes a 2D FA\footnote{For dynamic UAV networks, this paper adopts the pixel-array antennas to implement FAs, owing to their low fabrication cost, compact form factor, low switching power consumption, and microsecond-level switching delay.} with the size of $ W_{\text{tx}} \lambda = W_1^{\text{tx}}\lambda \times W_2^{\text{tx}}\lambda$ and $ N_{\text{tx}} = N_1^{\text{tx}} \times N_2^{\text{tx}} $ ports, where $\lambda$ is the carrier wavelength. Accordingly, each user has a 1D FA with the size of $ W_{\text{rx}}\lambda$ and $ N_{\text{rx}} $ ports.

We assume the UAV flies from an initial position to a destination within a period of $S$ seconds, which is divided into $T$ slots with equal duration $\tau = \frac{S}{T}$. Given a time slot $t \in [1,T]$, the positions of the UAV and user~$k$s are defined as $\bm{q}(t) = [x_q(t), y_q(t), H]$ and $\bm{p}^{\text{rx}}_k = [x_k, y_k,0]$, respectively. The velocity of the UAV is $v(t) = \frac{1}{\tau}||\bm{q}(t) - \bm{q}(t-1)||$, where $||\cdot||$ denotes the 2-norm of a vector. The yaw of the UAV in the direction of the motion is $\alpha(t)$. 
The position of $n$-th transmit FA ports is defined as $\hat{\bm{p}}_n^{\text{tx}} = [\hat{x}_n^{\text{tx}}, \hat{y}_n^{\text{tx}}]^\text{T}$ in a local coordinate system centered at the UAV origin, with $\hat{x}_n^{\text{tx}} \in [0, W_1^{\text{tx}}\lambda]$, $\hat{y}_n^{\text{tx}} \in [0, W_2^{\text{tx}}\lambda]$, where $[\cdot]^{\text{T}}$ denotes the transpose operation.
Similarly, the position of $n$-th receive FA port of user~$k$ is represented by $\hat{x}_n^{\text{rx}}$ in a local linear region centered at $\bm{p}^{\text{rx}}_k$, with $\hat{x}_n^{\text{rx}} \in [0, W^{\text{rx}}\lambda]$. We assume that the transmit FAs and receive FAs activated $L_{\text{tx}} \leq N_{\text{tx}}$ and $L_{\text{rx}} \leq N_{\text{rx}}$ ports, respectively. The channel\footnote{This paper assumes that the large-scale fading, angles of arrival/departure, and UAV attitude information can be obtained through existing channel estimation techniques. Thus the assumption of perfect CSI is adopted.} from the UAV to the user~$k$ can be represented by
\begin{equation}
\small
        \bm{H}_k(t) = PL_k(t)\left( Pr_{\text{LoS},k}(t) \bm{h}_{\text{LoS},k}(t) + Pr_{\text{NLoS},k}(t) \bm{h}_{\text{NLoS},k}(t) \right).
\end{equation}
Herein, the large-scale path loss (in dB) is
\begin{equation}
\small
  PL_k(t) = 20\log_{10}\!\left(\frac{c}{\lambda}\right) + 20\log_{10}\!\left(\frac{4\pi}{c}\right) + 20\log_{10}\!\big(d_k(t)\big).
  \label{eq:PL}
\end{equation}
where $c$ denotes the speed of light, $d_k(t) = \|\bm{q}(t) - \bm{p}^{\mathrm{rx}}_k\|$ denotes the UAV-to-user distance. The LoS path probability follows the empirical A2G model~\cite{nn1}
\begin{equation}
\small
  Pr_{\mathrm{LoS},k}(t) = \frac{1}{1 + a\exp\!\left(-b\!\left(\theta_k(t) - a\right)\right)},
  \label{eq:Plos}
\end{equation}
where $a$ and $b$ are environment-dependent constants, $\theta_k(t) = \arcsin(H/d_k(t))$ is the elevation angle between the UAV and the user $k$. Correspondingly, $Pr_{\mathrm{NLoS},k}(t) = 1 - Pr_{\mathrm{LoS},k}(t)$ is the Non-Line of Sight (NLoS)  probability. $\bm{h}_{\mathrm{LoS},k}(t),\, \bm{h}_{\mathrm{NLoS},k}(t) \in \mathbb{C}^{N_{\mathrm{rx}} \times N_{\mathrm{tx}}}$ denote the complex channel response of LoS and NLoS path, respectively. The LoS channel response is expressed as
\begin{equation}
\small
    \bm{h}_{\mathrm{LoS},k}(t) = \widetilde{D}_k(t) (\bm{S}^{\text{rx}}_k(t))^{\text{T}} \bm{a}^{\text{rx}}_k(t) (\bm{a}^{\text{tx}}_k(t))^{\text{H}} \bm{S}_\text{tx}(t),
\end{equation}
where $(\cdot)^{\mathrm{H}}$ denotes the conjugate transpose. The dynamic Doppler shift caused by the UAV motion is
\begin{equation}
\small
  \widetilde{D}_k(t) = \exp\!\left(j\frac{2\pi}{\lambda}\,v(t)\cdot t \cdot \cos\!\big(\phi_k(t)-\alpha(t)\big)\cos\theta_k(t)\right),
  \label{eq:Doppler}
\end{equation}
where $\phi_k(t) = \arctan(y_k - y_q(t)/x_k - x_q(t))$ is the azimuth angle between the UAV to the user $k$. $\bm{S}^{\mathrm{rx}}_k(t)\in\{0,1\}^{N_{\mathrm{rx}}\times L_{\mathrm{rx}}}$ and $\bm{S}_{\mathrm{tx}}(t)\in\{0,1\}^{N_{\mathrm{tx}}\times L_{\mathrm{tx}}}$ denote the receive and transmit port selection matrices, respectively. Constrained by the hardware, port selection matrices should satisfy
\begin{subequations}
\small
\begin{align}
&||\bm{S}_{\text{tx}}(t)(i,:)||\leq 1,\
||\bm{S}_{\text{tx}}(t)(:,j)|| = 1, \ \forall i\in [N_\text{tx}], j \in [L_\text{tx}], t,  \label{3a}\\
&||\bm{S}^{\text{rx}}_k(t)(i,:)||\leq 1,\
||\bm{S}^{\text{rx}}_k(t)(:,j)|| = 1, \ \forall i\in [N_\text{rx}],j \in [L_\text{rx}], k,t. \label{3b}
\end{align}
\end{subequations}
The 1D receive and 2D transmit FA phase-response vectors can be calculated as
\begin{subequations}
\small
\begin{align}
&\bm{a}^{\mathrm{rx}}_k(t) = \left[e^{j\frac{2\pi}{\lambda}\cos\theta_k(t)\,\hat{x}_1^{\mathrm{rx}}},\, \ldots,\, e^{j\frac{2\pi}{\lambda}\cos\theta_k(t)\,\hat{x}_{N_{\mathrm{rx}}}^{\mathrm{rx}}}\right]^{\!\mathrm{T}}, \label{eq:arx}\\
&\bm{a}^{\mathrm{tx}}_k(t) = \left[e^{j\frac{2\pi}{\lambda}\bm{k}_{\mathrm{tx}}^{\mathrm{T}}(t)\hat{\bm{p}}_1^{\mathrm{tx}}},\, \ldots,\, e^{j\frac{2\pi}{\lambda}\bm{k}_{\mathrm{tx}}^{\mathrm{T}}(t)\hat{\bm{p}}_{N_{\mathrm{tx}}}^{\mathrm{tx}}}\right]^{\!\mathrm{T}},
  \label{eq:atx}
\end{align}
\end{subequations}
where $ \bm{k}_\text{tx}(t) = [ \text{sin}\theta_k(t) \text{cos}(\phi_k(t)-\alpha(t)), \text{sin}\theta_k(t)\text{sin}(\phi_k(t)-\alpha(t))]$ is the array steering vector. The NLoS channel response is expressed as
\begin{equation}
\small
    \bm{h}_{\text{NLoS},k}(t) = \widetilde{D}_k(t)(\bm{S}^{\text{rx}}_k(t))^{\text{T}}\widetilde{\bm{H}}(t)\bm{S}_\text{tx}(t),
\end{equation}
where $\widetilde{\bm{H}}(t)$ is the spatial correlated channels of FAs following the definition in \cite{7}. Therefore, the receive Signal to Interference plus Noise Ratio (SINR) of user $k$ is
\begin{equation}
\small
\gamma_{k}(t)
=\frac{\left|\left(\bm{w}^{\text{rx}}_{k}(t)\right)^\text{H}\bm{H}_{k}(t)\bm{w}^{\text{tx}}_{k}(t)\right|^{2}}{\sum^K_{k^{\prime}\neq k}\left|\left(\bm{w}^{\text{rx}}_{k}(t)\right)^\text{H}\bm{H}_{k}(t)\bm{w}^{\text{tx}}_{k^{\prime}}(t)\right|^{2}+N_{0}},
\end{equation}
where $N_0$ denotes the noise power at user~$k$, $\bm{w}^{\text{rx}}_{k}(t)$, and $\bm{w}^{\text{tx}}_{k}(t)$ are the receive/transmit beamforming of user $k$, respectively.. The achievable data rate of user~$k$ at time slot $t$ is given by $R_k(t) = \text{log}_2(1 + \gamma_{k}(t))$.

In this work, we intend to maximize the average achievable data rate of all users with the arguments of UAV trajectory $\bm{Q} \triangleq \{\bm{q}(t), \forall t\}$, transmit FA port selection matrix $\bm{\mathcal{S}}_{\text{tx}} \triangleq \{\bm{S}_{\text{tx}}(t), \forall t\}$, transmit beamforming $\bm{W}_{\text{tx}} \triangleq \{\bm{w}_k^{\text{tx}}(t), \forall k,t\}$, receive FA port selection matrix $\bm{\mathcal{S}}_{\text{rx}} \triangleq \{\bm{S}^{\text{rx}}_k(t), \forall k,t\}$, receive beamforming $\bm{W}_{\text{rx}} \triangleq \{\bm{w}_k^{\text{rx}}(t), \forall k,t\}$. Therefore, the problem is formulated as
\vspace{-1ex}
\begin{subequations}
\small
\begin{align}
(\textbf{P1}): 
\quad\max_{\substack{ \bm{Q},\,\bm{\mathcal{S}}_{\text{tx}},\,\bm{W}_{\text{tx}} \\
\bm{\mathcal{S}}_{\text{rx}},\,\bm{W}_{\text{rx}}}}
& \frac{1}{T} \sum_{t=1}^T \sum_{k=1}^K R_k(t)  \label{6a}\\
\text{s.t.}\quad
& (\text{\ref{3a}}), (\text{\ref{3b}}),\, \bm{q}(T) = \bm{q}_\text{final}, \label{6b}\\
& ||\bm{q}(t) - \bm{q}(t-1) || \leq V_{\max} \tau, \ \forall t, \label{6c}\\
& \sum_{k=1}^K \| \bm{w}^{\text{tx}}_k(t) \|^2 \leq P_{\max}, \ \forall t. \label{6d}
\end{align}
\end{subequations}
Constraints (\ref{3a}) and (\ref{3b}) denote the FA deployment limitations, (\ref{6b}) and (\ref{6c}) are the UAV motion limitations, (\ref{6d}) represents the transmit power limitation.
\vspace{-1ex}
\section{Proposed AO Algorithm}

Problem (\textbf{P1}) involves five coupled optimization blocks: the UAV trajectory $\bm{Q}$, the transmit FA port selection $\bm{\mathcal{S}}_{\mathrm{tx}}$, the transmit beamforming $\bm{W}_{\mathrm{tx}}$, the receive FA port selection $\bm{\mathcal{S}}_{\mathrm{rx}}$, and the receive beamforming $\bm{W}_{\mathrm{rx}}$. The joint optimization is highly non-convex due to: (i) the fractional SINR structure coupling all blocks; (ii) the combinatorial binary FA port selection constraints; and (iii) the rank-one beamforming constraints, resulting an NP-hard global optimization.

Directly obtaining the global optimum of the non-convex problem is highly challenging and cannot be achieved in a polynomial time. In contrast, obtaining a convergent suboptimal solution via AO is more tractable. Specifically, the AO method fixes part of the variables in each iteration and alternately optimizes the remaining variable blocks, to gradually improve the objective value.
\vspace{-2ex}
\subsection{UAV Trajectory Optimization}
Given the $\{\bm{\mathcal{S}}_{\text{tx}}, \bm{W}_{\text{tx}}, \bm{\mathcal{S}}_{\text{rx}}, \bm{W}_{\text{rx}}\}$, we aim to optimize the UAV trajectory $\bm{Q}$. As $\bm{a}^{\text{tx}}_k(t)$, $\bm{a}^{\text{rx}}_k(t)$, $\widetilde{D}_k(t)$, and $Pr_{\text{LoS},k}(t)$ exhibit complexity and non-linearity with respect to $\bm{q}(t)$, we utilize the $i$-th iteration of the UAV trajectory $\bm{q}^{(i)}(t)$ to approximate these variables. For each user $k$ and time slot $t$, we employ $\{C_{k,j}(t)\}_{j=1}^K$ to represent the effective channel gains
\begin{align}
\small
    C_{k,j}(t) & = \Big(\left( \bm{w}^{\text{rx}}_k(t)\right)^{\text{H}} \big( Pr_{\text{LoS},k}(t) \bm{h}_{\text{LoS},k}(t) \\ 
    & + Pr_{\text{NLoS},k}(t) \bm{h}_{\text{NLoS},k}(t) \big) \bm{w}^{\text{tx}}_j(t)\Big)^2.
    \notag
\end{align}
Then the rate formula of user $k$ at time $t$ can be rewritten as
\begin{align}
\small
    R_k\left(\bm{q}(t)\right) & = \log_2\Big(\frac{\sum_{j=1}^K C_{k,j}(t)}{(d_k(t))^2}+N_0\Big) \\
    &- \log_2\Big(\frac{\sum_{j\ne k}^K C_{k,j}(t)}{(d_k(t))^2}+N_0\Big). \notag
\end{align}
Then we introduce an auxiliary variable $p_{k}(t)\ge 0$ to represent the distance-related term
\begin{align}
\small
    &f_{k}(t) = \log_2\Big(p_{k}(t)\sum_{j=1}^K C_{k,j}(t)+N_0\Big), \\
    &z_{k}(t) = \log_2\Big(p_{k}(t)\sum_{j\ne k} C_{k,j}(t)+N_0\Big).
\end{align}
However, the function $f_{k}(t) - z_{k}(t)$ is still non-convex, thus we further apply Successive Convex Approximation (SCA) with the first-order Taylor expansion at the $i$-th iteration of $p_{k}^{(i)}(t)$ to obtain the upper bound for $z_k(t)$. Therefore, (\textbf{P1}) can be reformulated as
\vspace{-2ex}
\begin{subequations}
\small
\begin{align}
(\textbf{P2}): 
& \quad\max_{\substack{ \bm{Q},\, \bm{P} = \{ p_k(t), \forall k,t\}}}
\frac{1}{T}\sum_{t=1}^T \sum_{k=1}^K f_k(t) - \hat z_{k}(t) \label{9a}\\
\text{s.t.}\quad
& \bm{q}(T) = \bm{q}_\text{final}, \label{9b}\\
& ||\bm{q}(t) - \bm{q}(t-1) || \leq V_{\max} \tau, \, \forall t, \label{9c}\\
& ||\bm{q}(t) - \bm{p}_\text{rx,k} ||^2 \leq \frac{1}{p_{k}^{(i)}(t)} - \frac{1}{p_{k}^{(i)2}(t)}\left(p_k(t) - p_{k}^{(i)}(t)\right),\, \forall t,
\end{align}
\label{P2}
\end{subequations}
\vspace{-1ex}

\noindent where $\hat{z}_k(t) = z_k\!\left(p_k^{(i)}(t)\right) + \nabla z_k\!\left(p_k^{(i)}(t)\right)\!\left(p_k(t) - p_k^{(i)}(t)\right)$ and $\nabla z_{k}\left(p_{k}^{(i)}(t)\right) = \frac{\sum_{j\ne k}^K C_{k,j}(t)} {(p_{k}^{(i)}(t)\sum_{j\ne k} C_{k,j}(t) + N_0)\cdot \text{ln}2}$.
Therefore, Problem (\textbf{P2}) is a Second-Order Cone Programming (SOCP) problem and can be solved by standard convex solvers.

\begin{remark}[Convergence of (\textbf{P2})]
The SCA surrogate $\hat{z}_k(t)$ satisfies two conditions: (a) $\hat{z}_k(t)\big|_{p_k = p_k^{(i)}} = z_k\!\left(p_k^{(i)}\right)$; (b) $\nabla \hat{z}_k\big|_{p_k=p_k^{(i)}} = \nabla z_k\!\left(p_k^{(i)}\right)$. Since $z_k(p_k)$ has $\frac{d^2z_k}{dp_k^2}=-\frac{B_k^2}{(p_kB_k+N_0)^2\ln2}<0$, it is concave for $p_k$. The first-order Taylor expansion of $z_k(p_k)$ is an upper bound: $z_k(p_k)\leq\hat{z}_k(p_k)$ for all $p_k \ge 0$. Consequently, the objective is the lower bound of $R_k(\bm{q}(t))$: $f_k(t)-\hat{z}_k(t)\leq f_k(t)-z_k(t)$. The monotone non-decrease follows the chain rule
\begin{equation*}
  R_k^{(i)} = f_k^{(i)}-\hat{z}_k^{(i)}\big|_{p_k=p_k^{(i)}} \leq f_k^{(i+1)}-\hat{z}_k^{(i+1)}\big|_{p_k=p_k^{(i+1)}} \leq R_k^{(i+1)}.
\end{equation*}
Given that the lower bound of the total achievable data rate is non-decreasing over iterations and has an upper bound, (\textbf{P2}) is ensured to converge.
\end{remark}

\vspace{-2ex}

\subsection{Transmit FA Port Selection Matrix Optimization}
Given the $\{\bm{Q}, \bm{W}_{\text{tx}}, \bm{\mathcal{S}}_{\text{rx}}, \bm{W}_{\text{rx}}\}$, we aim to optimize the FA port selection matrix $\bm{\mathcal{S}}_{\text{tx}}$. Firstly, we reformulate the objective function of (\textbf{P1}) for each time slot $t$ as
\begin{equation}
\begin{aligned}
\small
    R_k\left(\bm{S}_{\text{tx}}^{\text{T}}(t)\right)  & = \log_2\left(\text{Tr}\left(\bm{S}_{\text{tx}}^{\text{T}}(t) \bm{A}_k(t) \bm{S}_{\text{tx}}(t) \bm{D}_{\text{all}}\right) + N_0 \right) \\ 
    & - \log_2 \left(\text{Tr}\left(\bm{S}_{\text{tx}}^{\text{T}}(t) \bm{A}_k(t) \bm{S}_{\text{tx}}(t) \bm{D}_{\text{int}}\right) + N_0\right),
\end{aligned}
\end{equation}
where $\text{Tr}\{\cdot\}$ denotes the trace of the matrix, $\bm{D}_{\text{all}}(t) = \sum_{j=1}^{K} \bm{w}^{\text{tx}}_{j}(t) (\bm{w}^{\text{tx}}_{j}(t))^{\text{H}}$, and $ \bm{D}_{\text{int}}(t) = \sum_{j \neq k}^{K} \bm{w}^{\text{tx}}_{j}(t) (\bm{w}^{\text{tx}}_{j}(t))^{\text{H}}$. We introduce auxiliary variables $\bm{A}_k(t) = \bm{C}_k^{\text{H}}(t) \bm{C}_k(t)$, where 
\begin{equation}
\small
    \bm{C}_k(t) = PL_k(t) \left( \bm{w}_k^{\text{rx}}(t) \right)^{\text{H}} (\bm{S}^{\text{rx}}_k (t))^{\text{T}}  \widetilde{D}_k(t) \hat{\bm{H}}_k(t),
\end{equation}
and
\begin{equation}
\small
    \hat{\bm{H}}_k(t) = Pr_{\text{LoS},k}(t) \bm{a}^{\text{rx}}_k(t) (\bm{a}^{\text{tx}}_k(t))^{\text{H}} + Pr_{\text{NLoS},k}(t) \widetilde{\bm{H}}(t).
\end{equation}
To handle the non-convexity, we adopt SCA at the $i$-th FA port selection iterate $\bm{S}_{\mathrm{tx}}^{(i)}(t)$. $R_k(t)$ can be reformulated as
\begin{equation}
\small
  \hat{R}_k(t) = R_k\!\left(\bm{S}_{\mathrm{tx}}^{(i)}(t)\right) + \mathrm{Tr}\!\left(\nabla R_k^{\mathrm{T}}\!\left(\bm{S}_{\mathrm{tx}}^{(i)}(t)\right)\!\left(\bm{S}_{\mathrm{tx}}(t) - \bm{S}_{\mathrm{tx}}^{(i)}(t)\right)\right),
  \label{eq:Rhat}
\end{equation}
where
\begin{equation}
\small
  \nabla R_k\!\left(\bm{S}_{\mathrm{tx}}^{(i)}(t)\right) = \frac{2}{\ln 2}\,\bm{A}_k(t)\,\bm{S}_{\mathrm{tx}}^{(i)}(t)\!\left(\frac{\bm{D}_{\mathrm{all}}(t)}{s_{\mathrm{all}}(t)} - \frac{\bm{D}_{\mathrm{int}}(t)}{s_{\mathrm{int}}(t)}\right),
  \label{eq:gradR}
\end{equation}
with the denominators calculated at the each iteration
\begin{align}
\small
  s_{\mathrm{all}}(t) &= \mathrm{Tr}\!\left(\!\left(\bm{S}_{\mathrm{tx}}^{(i)}\right)^{\!\mathrm{T}}\!\bm{A}_k(t)\,\bm{S}_{\mathrm{tx}}^{(i)}\bm{D}_{\mathrm{all}}\right) + N_0, \label{eq:sall}\\
  s_{\mathrm{int}}(t) &= \mathrm{Tr}\!\left(\!\left(\bm{S}_{\mathrm{tx}}^{(i)}\right)^{\!\mathrm{T}}\!\bm{A}_k(t)\,\bm{S}_{\mathrm{tx}}^{(i)}\bm{D}_{\mathrm{int}}\right) + N_0. \label{eq:sint}
\end{align}
Since the subproblem for $\bm{S}_{\mathrm{tx}}(t)$ is independent across time slots, (\textbf{P1}) decomposes into $T$ per-slot subproblems. Each is reformulated as:
\vspace{-1.5ex}
\begin{equation}
(\textbf{P3}): \max_{\bm{S}_{\text{tx}}(t)} \sum_{k=1}^{K} \text{Tr}\left(\nabla R^{\text{T}}_k(\bm{S}_{\text{tx}}^{(i)}(t))\bm{S}_{\text{tx}}(t)\right)
\quad \text{s.t.} (\text{\ref{3a}}), (\text{\ref{3b}}).
\label{P3}
\end{equation}
To this end, (\textbf{P3}) is a Mixed-Integer Linear Programming (MILP) problem and can be solved by CVX toolbox.

\begin{remark}[Convergence of (\textbf{P3})]
The linear programming relaxation of the binary constraints~(\ref{3a}) and (\ref{3b}) yields a standard assignment polytope whose constraint matrix is totally unimodular. The optimal solution is always an integer vertex, ensuring binary port selection. Moreover, since $R_k\!\left(\bm{S}_{\mathrm{tx}}^{\mathrm{T}}\right)$ is concave in $\bm{S}_{\mathrm{tx}}$, the first-order surrogate $\hat{R}_k(t)$ satisfies $\hat{R}_k(t) \leq R_k\!\left(\bm{S}_{\mathrm{tx}}^{\mathrm{T}}\right)$ globally, confirming that the MILP objective is monotonically non-decreasing across iterations.
\end{remark}

\vspace{-2ex}
\subsection{Transmit FA Beamforming Optimization}
Given the $\{\bm{Q}, \bm{\mathcal{S}}_{\text{tx}}, \bm{\mathcal{S}}_{\text{rx}}, \bm{W}_{\text{rx}}\}$, we aim to optimize the transmit FA beamforming $\bm{W}_{\text{tx}}$. For each time slot $t$, we first introduce the auxiliary variables $\bm{F}(t)=\{\bm{F}_k(t)\}_{k=1}^{K}$, where $\bm{F}_k(t) = \bm{w}^{\text{tx}}_k(t)\big(\bm{w}^{\text{tx}}_k(t)\big)^{H}$. Note that $\bm{F}_k(t)$ is semi-definite, i.e. $\bm{F}_k(t) \succeq 0$, with the rank-one constraint. Accordingly, the rate of user $k$ at time slot $t$ can be rewritten as $R_k(\bm{F}(t)) = f_k(\bm{F}(t)) - z_k(\bm{F}(t))$, where
\begin{align}
\small
    & f_k\left(\bm{F} (t)\right) = \log_2 \Big( \sum_{j=1}^{K}\text{Tr}\big(\bm{F}_j(t)\bm{G}_k(t)\big)+N_0\Big), \\
    & z_k\left(\bm{F}(t)\right) = \log_2 \Big(\sum^K_{j\ne k}\text{Tr}\big(\bm{F}_j(t)\bm{G}_k(t)\big)+N_0\Big).
\end{align}
$\bm{G}_k(t)= \bm{g}_k(t)\bm{g}_k(t)^{H}$ is the effective channel matrix, where $\bm{g}_k(t) = \big(\bm{w}^{\text{rx}}_k(t)\big)^{\text{H}}\bm{H}_{k}(t)\in\mathbb{C}^{L_{\text{tx}}\times 1}$. However, $f_{k}(\bm{F}(t)) - z_{k}(\bm{F}(t))$ is still non-convex. Thus, we employ SCA at the $i$-th iteration of the $\bm{F}^{(i)}(t)$ to obtain the upper bound of $z_k(\bm{F}(t))$
\begin{equation}
\small
  \tilde{z}_k\!\left(\bm{F}(t)\right)\! = \!z_k\!\left(\bm{F}^{(i)}(t)\right) \!+\! \sum_{j\neq k}^{K}\!\left\langle\nabla_{\!\bm{F}_j} z_k\!\left(\bm{F}_j^{(i)}(t)\right),\bm{F}_j(t) \!-\! \bm{F}_j^{(i)}(t)\right\rangle,
  \label{eq:zhat_F}
\end{equation}
where $\langle\bm{A},\bm{B}\rangle = \text{Re}\{\text{Tr}(\bm{A}^{\mathrm{H}}\bm{B})\}$ and $
  \nabla_{\!\bm{F}_j} z_k\!\left(\bm{F}_j^{(i)}(t)\right) = \frac{\bm{G}_k(t)}{\!\left(\sum_{j'\neq k}\mathrm{Tr}\!\left(\bm{F}_{j'}^{(i)}(t)\bm{G}_k(t)\right)+N_0\right)\!\ln 2}$.
To handle the rank-one constraints of $\bm{F}_k(t)$, we utilize the Sequential Rank-One Constraint Relaxation (SROCR) method \cite{8}. At iteration $i$, the rank-one constraint is gradually imposed by
\begin{equation}
\small
\big(\bm{\lambda}_k^{(i)}\big)^{H}\bm{F}_k(t)\bm{\lambda}_k^{(i)}
\ge \tau_k^{(i)}\,\text{Tr}\big(\bm{F}_k(t)\big),\tau_k^{(i)} \in [0, 1],\ \forall k,
\label{srocr}
\end{equation}
where $\bm{\lambda}_k^{(i)}$ denote the eigenvector of $\bm{F}_k^{(i)}(t)$, $\tau_k^{(i)}$ is a relaxation parameter updated progressively towards $1$. Note that the subproblem for $\bm{F}(t)$ is independent across the time slots, the problem (\textbf{P1}) can be decomposed and reformulated as
\vspace{-1.5ex}
\begin{subequations}
\small
\begin{align}
(\textbf{P4}): 
& \quad\max_{\substack{\bm{F}(t)}}
\sum_{k=1}^K f_k(\bm{F}(t)) - \tilde z_{k}(\bm{F}(t)) \\
\text{s.t.}\quad
& \bm{F}_k(t) \succeq 0,\, \sum_{k=1}^{K}\text{Tr}\big(\bm{F}_k(t)\big)\le P_{\max},\ (\ref{srocr}), \ \forall t.
\end{align}
\label{P4}
\end{subequations}
\vspace{-2ex}

\noindent Consequently, (\textbf{P4}) is a semidefinite programming problem which can be solved by CVX toolbox. After convergence, $\bm{w}^k_{\text{tx}}(t)$ can be recovered by eigen-decomposition.

\begin{remark}[Convergence of \textbf{P4}]
The surrogate $\tilde {z}_k(\bm{F}(t))$ satisfies $z_k(\bm{F}(t)) \geq \hat{z}_k(\bm{F}(t))$, so $f_k - \hat{z}_k$ is a valid lower bound on $R_k$. Besides, the SROCR constraint~\eqref{srocr} is convex for fixed $\bm{\lambda}_k^{(i)}$ and becomes convergent as $\tau_k^{(i)}\to 1$, progressively enforcing rank-one structure. Therefore, the current iterate $\bm{F}_k^{(i)}(t)$ satisfies all constraints, guaranteeing feasibility throughout. Based the above analysis, the objective is non-decreasing across iterations.
\end{remark}

\vspace{-2ex}
\subsection{Receive FA Port Selection Matrix Optimization}
Given the $\{\bm{Q}, \bm{\mathcal{S}}_{\text{tx}}, \bm{W}_{\text{tx}}, \bm{W}_{\text{rx}}\}$, we aim to optimize the receive FA port selection matrix $\bm{\mathcal{S}}_{\text{rx}}$ through Fluid Antenna Multiple Access (FAMA)~\cite{9}. Specifically, we first obtain the effective channel matrices of user $k$ at time slot $t$
\begin{equation}
\bm{D}_k(t) = PL_k(t) \widetilde{D}_k(t) \hat{\bm{H}}(t)\bm{S}_{\text{tx}}(t)
\in\mathbb{C}^{N_{\text{rx}}\times L_{\text{tx}}}.
\end{equation}
For each receive candidate port $n\in\{1,\ldots,N_{\text{{rx}}}\}$, the desired signal power of user $k$ is $
P^{\text{sig}}_{k,n}(t)=\left|\big[\bm{D}_k(t) \bm{w}^{\text{tx}}_k(t)\big]_n\right|^2
$, where $[\cdot]_n$ denote the $n$-th element of a vector, while the multi-user interference power is
$
P^{\text{int}}_{k,n}(t)=\sum_{j\ne k}\left|\big[\bm{D}_k(t) \bm{w}^{\text{tx}}_j(t)\big]_n\right|^2.
$
Accordingly, the SINR at $n$-th port of user $k$ is given by
$
\text{SINR}_{k,n}(t)=\frac{P^{\text{sig}}_{k,n}(t)}{P^{\text{int}}_{k,n}(t) + N_0},
$
Finally, we activate the top $L_{\text{rx}}$ ports of user $k$ with the largest SINR in each time slot $t$
\begin{equation}
    l^*_{k,n} = \underset{l \in \{N_{\text{rx}}\} \setminus \{l^* _{k,1}, \dots, l^*_{k,n-1}\}}{\arg\max}\ \text{SINR}_{k,n}(t), \, l^*_{k,n} \in [0, L_{\text{rx}}].
\label{P5}
\end{equation} 
Given the index set of the activated ports, we can construct $\bm{S}_{\text{rx},k}(t)$ by selecting the corresponding matrix columns.

\vspace{-2ex}
\subsection{Receive FA Beamforming Optimization}
Given the $\{\bm{Q}, \bm{W}_{\text{tx}}, \bm{\mathcal{S}}_{\text{tx}}, \bm{\mathcal{S}}_{\text{rx}} \}$, we aim to optimize the receive FA beamforming $\bm{W}_{\text{rx}}$ through the Minimum Mean Square Error (MMSE) receiver for user $k$ at time slot $t$
\vspace{-1ex}
\begin{equation}
\bm{w}^{\text{rx}}_k(t)= \left( N_0 \bm{I}_{L_{\text{rx}}} + \sum_{j=1}^{K}\bm{R}_j(t)\big( \bm{R}_j(t)\big)^{\text{H}} \right)^{-1} \bm{R}_k(t),
\label{P6}
\end{equation}
where $\bm{I}_{L_{\text{rx}}}$ denotes the identity matrix of size $L_{\text{rx}} \times L_{\text{rx}}$ and $\bm{R}_k(t) = \bm{H}_k(t) \bm{w}^{\text{tx}}_k(t)$.

\subsection{Overall AO Algorithm Analysis}
The overall algorithm can be conducted by iteratively solving $\bm{Q}$ in (\ref{P2}), solving $\bm{\mathcal{S}}_{\text{tx}}$ in (\ref{P3}), solving $\bm{W}_{\text{tx}}$ in (\ref{P4}), solving $\bm{\mathcal{S}}_{\text{rx}}$ in (\ref{P5}), solving $\bm{W}_{\text{rx}}$ in (\ref{P6}). The convergence and complexity analysis of the algorithm are given below.

\subsubsection{Convergence of the AO Algorithm}
Remark~1,  Remark~2, and Remark~3 guarantee the (\textbf{P2}), (\textbf{P3}), (\textbf{P4}) solution is non-decreasing. The greedy SINR maximization~\eqref{P5} globally maximizes per-user SINR for fixed transmit variables, so $f$ cannot decrease. Finally, the closed-form MMSE solution~\eqref{P6} uniquely minimizes MSE for fixed transmit variables, so $f$ cannot decrease. Therefore, the lower bound of total achievable rate is monotonically non-decreasing and converges to a finite limit. Moreover, the objective has an upper bounded under the finite power constraint $P_{\max}$ and bounded antenna apertures. Then the AO Algorithm is ensure to converge and get suboptimal solutions.

\begin{figure}[!t]
    \centering
    \subfloat[]{
        \includegraphics[width=0.8\linewidth]{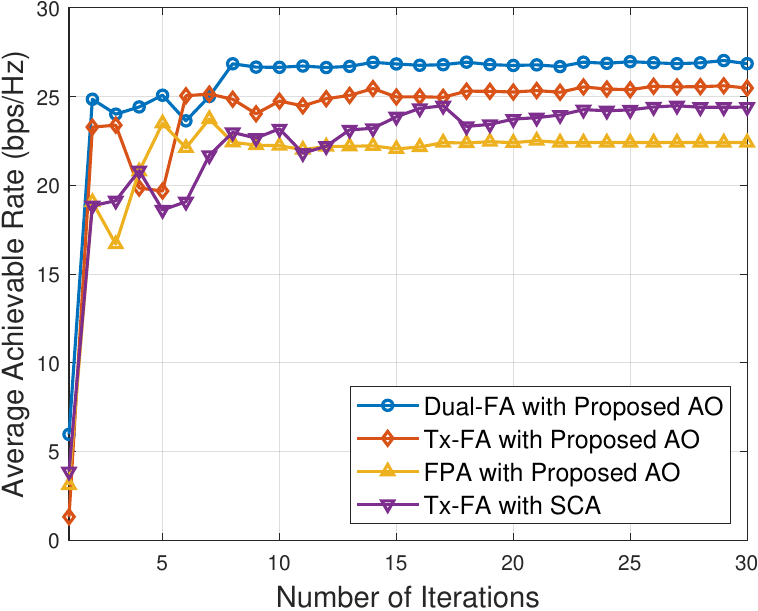}
        \label{fig:convergence}
    }
    \hfill
    \subfloat[]{
        \includegraphics[width=0.8\linewidth]{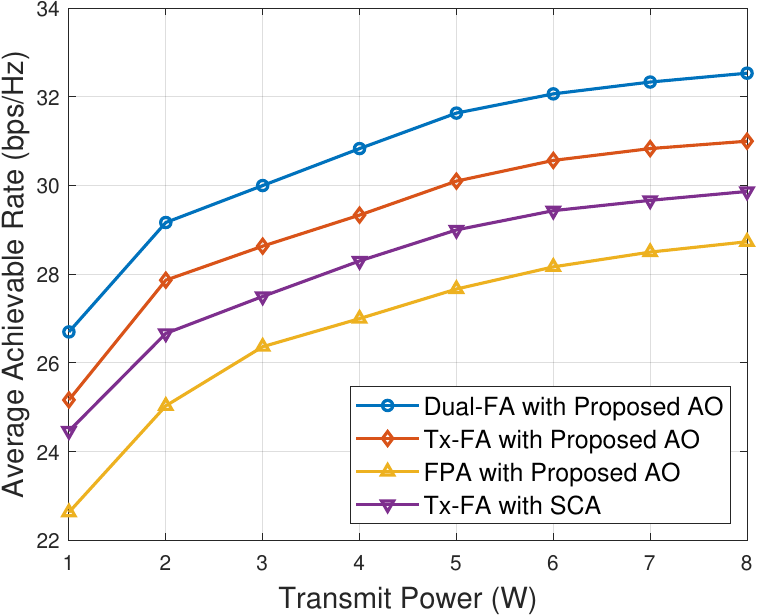}
        \label{fig:power}
    }
    \caption{(a) Convergence of the AO algorithm. (b) Average achievable data rate versus transmit power.}
    \label{fig:row1}
    \vspace{-3ex}
\end{figure}

\subsubsection{Complexity of the AO Algorithm}
The complexity of the proposed AO algorithm is analyzed as follows.
For the trajectory subproblem (\textbf{P2}), the dominant cost comes from solving an SOCP with $KT$ optimization variables (the 3-D UAV positions and auxiliary scalars $p_k(t)$) subject to $\mathcal{O}(KT)$ second-order cone constraints. Using interior-point methods, this incurs a complexity of $\mathcal{O}\!\left((KT)^3\right)$ per AO iteration.
For the transmit port selection subproblem (\textbf{P3}), the problem is decomposed into $T$ independent slots, each with $N_{\mathrm{tx}}L_{\mathrm{tx}}$ optimization variables corresponding to the $N_{\mathrm{tx}}\times L_{\mathrm{tx}}$ selection matrix. The total complexity is $\mathcal{O}\!\left(T(N_{\mathrm{tx}}L_{\mathrm{tx}})^3\right)$.
For the transmit beamforming subproblem (\textbf{P4}), the problem is similarly decomposed into $T$ independent slots. In each slot, there are $K$ Hermitian matrix variables of size $L_{\mathrm{tx}}\times L_{\mathrm{tx}}$, so the number of scalar unknowns is $KL_{\mathrm{tx}}^2$, and there are $K$ positive semidefinite constraints of dimension $L_{\mathrm{tx}}$. Hence the per-slot complexity is $\mathcal{O}(K^3L_{\mathrm{tx}}^6)$, giving a total of $\mathcal{O}\!\left(TK^3L_{\mathrm{tx}}^6\right)$.
For the receive port selection subproblem, the major cost is computing the per-port SINR values, which requires evaluating $N_{\mathrm{rx}}$ inner products of length $L_{\mathrm{tx}}$ per user per slot, followed by a sort. The total complexity is $\mathcal{O}\!\left(TKN_{\mathrm{rx}}\log_2 N_{\mathrm{rx}}\right)$.
For the receive beamforming subproblem, the MMSE solution requires inverting an $L_{\mathrm{rx}}\times L_{\mathrm{rx}}$ covariance matrix per user per slot, contributing $\mathcal{O}\!\left(TKL_{\mathrm{rx}}^3\right)$.
In summary, the overall complexity of the AO algorithm is
$
\mathcal{O}\big(\big[(TK)^3
+( TK^3L_{\text{tx}}^6 \big)
+(T(N_{\text{tx}}L_{\text{tx}})^3)
+(TKN_{\text{rx}}\text{log}_2{N_{\text{rx}}})
$
$
+(TK(L_{\text{rx}})^3)
] 
I_{\text{all}} \text{log}_2(\epsilon^{-1})\big)
$, where $I_{\text{all}}$ indicates the maximum number of iterations and $\epsilon$ is the precision threshold.
\vspace{-3ex}

\begin{figure}[!t]
    \centering
    \subfloat[]{
        \includegraphics[width=0.8\linewidth]{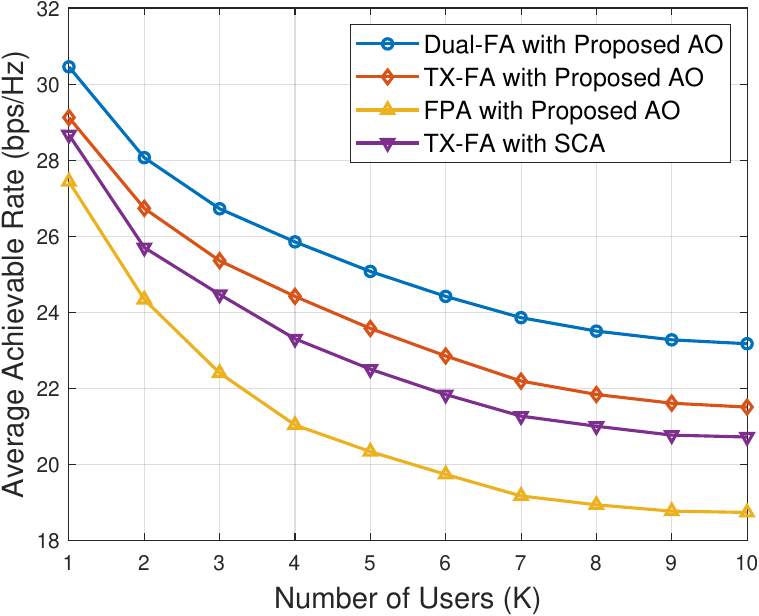}
        \label{fig:user}
    }
    \hfill
    \subfloat[]{
        \includegraphics[width=0.8\linewidth]{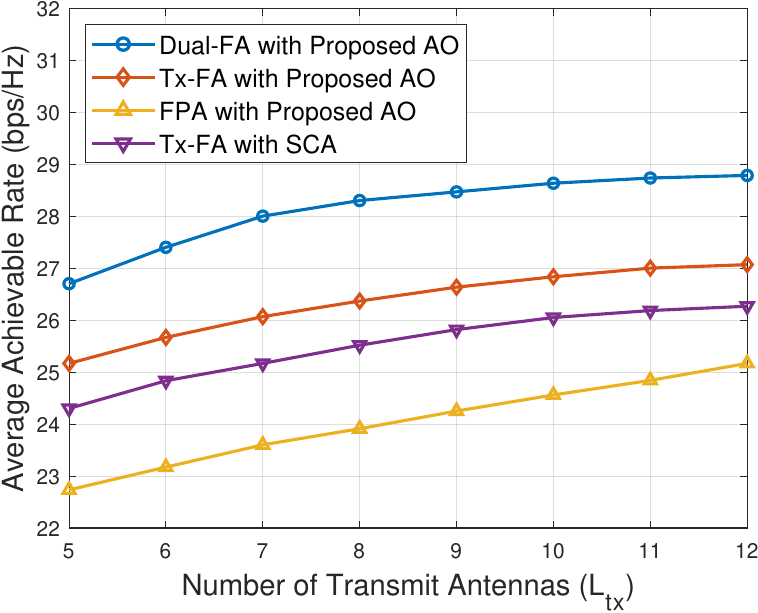}
        \label{fig:antenna}
    }
    \caption{Average achievable data rate versus (a) the number of users, (b) the number of transmit antennas.}
    \label{fig:row2}
\vspace{-3ex}
\end{figure}
\section{Numerical Results}
\setlength{\abovecaptionskip}{2pt} 
\setlength{\belowcaptionskip}{2pt} 

In this section, we provide numerical results to validate the superiority of the proposed AO algorithm. In the simulation, we set $a = 9.6, b = 0.16$, the area of $ 500\ \text{m} \times 500\ \text{m}$ with $K = 3$ users, the time period as $S = 30$ s, UAV flight altitude as $H = 120$ m, UAV maximum velocity as $V_{\text{max}} = 40$ m/s, the maximum transmit power as $P_{\text{max}} = 1 $ W, the noise power at each receiver as $N_0 = -110$ dBm, the carrier frequency as $f_c$ = 2 GHz, the 2D FA with $W_{\text{tx}} = 2\lambda \times 2\lambda$, $N_{\text{tx}} = 10 \times 10$ ports, and $L_\text{tx} = 5$ selected ports, the 1D FA with $W_{\text{rx}} = 2\lambda$, $N_{\text{rx}} = 10$ ports, and $L_\text{rx} = 3$ selected ports. We compare our proposed algorithm across three cases: 1) Tx-FA with proposed AO: the proposed AO is utilized, while the UAV is equipped with a 2D FA and each user is equipped with FPAs.
2) FPA with proposed AO: the proposed AO is utilized, while the UAV and each user are equipped with FPAs.
3) TX-FA with SCA: the SCA in \cite{6} is utilized, while the UAV is equipped with a 2D FA and each user is equipped with FPAs.

In Fig. \ref{fig:row1}(a), we illustrate the convergence of different algorithms. It can be observed that FA-based algorithms consistently outperform the FPA baseline, with the dual-FA achieving up to $120\% $ gains by exploiting more spatial DoFs. Moreover, the proposed algorithm attains faster convergence and higher rates than the SCA baseline, which requires nearly 20 iterations due to its relaxed rank-one beamforming constraint in dynamic environments.
Fig. \ref{fig:row1}(b) presents the average achievable rate versus the transmit power. As the transmit power increases, the proposed algorithm exhibits a larger performance gap compared with the FPA-based and SCA baselines, owing to its enhanced channel gains enabled by FA position adaptivity and flexible beamforming.
In Fig. \ref{fig:row2}(a), we illustrate the average achievable rate versus the number of users. The proposed algorithm manifests the effectiveness in MU interference mitigation over the FPA-based and SCA baselines, achieving $125\%$ and $112 \%$ performance gains, respectively.
Fig. \ref{fig:row2}(b) showcases the performance under different numbers of transmit antennas. By selecting favorable port combinations from additional spatial DoFs, the proposed algorithm achieves significant gains at the initial stage, while the benefits gradually saturate as the antenna count increases.
In general, the proposed dual-FA algorithm consistently exhibits superiority in throughput over other algorithms under the dynamic UAV networks.
\vspace{-1ex}
\section{Conclusion}
This letter investigated a dual FA-assisted UAV network for MU-MIMO downlink communications. We first introduced a general channel model to capture the phase-related features between mobile UAVs and users. To maximize the average achievable rate, we proposed an AO algorithm which iteratively optimizes the UAV trajectory, transmit/receive FA port selection, and beamforming for decomposition of the non-convex problems. Numerical results verified that the proposed algorithm achieves significant throughput gains over FPA-based and existing FA-based baselines. Moving forward, our research will extend to multi-UAV cooperation, imperfect channel information, and hardware-constrained FA designs.
\vspace{-2ex}

\end{document}